# RESONANCE ENERGY TRANSPORT IN AN OSCILLATOR CHAIN

Agnessa Kovaleva

Space Research Institute, Russian Academy of Sciences, Moscow 117997, Russia

We investigate energy transfer and localization in a linear time-invariant oscillator chain weakly coupled to a forced nonlinear actuator. Two types of perturbation are studied: (1) harmonic forcing with a constant frequency is applied to the actuator (the Duffing oscillator) with slowly changing parameters; (2) harmonic forcing with a slowly increasing frequency is applied to the nonlinear actuator with constant parameters. In both cases, stiffness of linear oscillators as well as linear coupling remains constant, and the system is initially engaged in resonance. The parameters of the systems and forcing are chosen to guarantee autoresonance (AR) with gradually increasing energy in the nonlinear actuator. As this paper demonstrates, forcing with constant frequency generates oscillations with growing energy in the linear chain but in the system excited by forcing with slowly time-dependent frequency energy remains localized on the nonlinear actuator whilst the response of the linear chain is bounded. This means that the systems that seem to be almost identical exhibit different dynamical behavior caused by their different resonance properties. Numerical examples a good agreement between exact (numerical) solutions and their asymptotic approximations found by the multiple time scales method.

## I. Introduction

It is well known that high-energy resonant oscillations in a linear oscillator can be generated by an external force whose constant frequency matches the frequency of the oscillator. In this case the amplitude of oscillations is defined by the constant forcing amplitude and frequency, and the change of the forcing and/or oscillator frequency is followed by escape from resonance. On the contrary, the frequency of a nonlinear oscillator changes as the amplitude changes, and the oscillator remains in resonance with its drive if the driving frequency and/or other parameters vary slowly in time to be consistent with the changing frequency of the oscillator. The ability of a nonlinear oscillator to stay captured into resonance due to variance of its structural or/and excitation parameters is known as autoresonance (AR).

After first studies for the purposes of particle acceleration [1-3], a large number of theoretical approaches, experimental results and applications of AR in different fields of natural science, from plasmas to planetary dynamics, have been reported in literature, see, e.g., [4-6] and references therein. The analysis was first concentrated on the study of AR in the basic nonlinear oscillator but then the developed methods and approaches were extended to two- or three-dimensional systems. Examples in this category are excitations of continuously phase-locked plasma waves with laser beams [7-9], particle transport in a weak external field with slowly changing frequency [10-12], control of vibrational and rotational degrees of freedom of diatomic molecules [13], etc.

In most of these studies, AR was considered as an effective method of excitation and control of high-energy oscillations in the entire system. In this work we demonstrate that this conclusion cannot be applied universally, because AR in the multi-dimensional system is a much more complicated phenomenon than AR in a single oscillator, and the behavior of each element in the multi-dimensional array may drastically differ from the dynamics of a single oscillator. These effects were recently analyzed for a two-degree-of-freedom (2DOF) cell consisting of weakly coupled linear and nonlinear oscillators [14]. This paper studies a more general problem of energy transfer and localization in a resonance multidimensional array consisting of a chain of time-invariant linear oscillators with equal partial frequencies weakly coupled to a nonlinear actuator (the Duffing oscillator) driven by an external force. Two types of excitations are considered: (1) an actuator with slowly time-decreasing linear stiffness is driven by a periodic force with constant frequency; (2) a time-independent nonlinear actuator is driven by a force with a slowly-increasing frequency. In both cases, the parameters of the linear chain and linear coupling remain constant, and the system is initially captured into resonance.

Since AR is a purely nonlinear effect, oscillations with growing energy in the linear chain may arise only in the presence of AR in the nonlinear actuator. It was shown [15] that the conditions of the occurrence of AR and the amplitude of oscillations in a single Duffing oscillator are equivalent for both types of excitation. The purpose of this paper is to find the conditions under which AR in the nonlinear actuator brings about growing oscillations in the coupled chain and define the parameters of possible dynamical regimes.

We recall that multidimensional nonlinear nonstationary systems seldom yield explicit analytical solutions needed for understanding and modelling the transition phenomena but numerical solutions are often insufficient for the comprehensive understanding of the underlying dynamics. In this paper the multiple scales formalism (see, e.g., [16]) is invoked to derive explicit asymptotic solutions for both types of dynamical systems.

The paper is organized as follows. Section II shows that, under appropriate conditions, periodic forcing with constant (resonant) frequency being applied to the nonlinear actuator with slowly-varying parameters gives rises to AR in both the actuator as well as in the coupled linear chain. The theoretical and numerical solutions indicate the equal energy distribution between all oscillators in the array. At the same time, as shown in Sec. III, a drive with a slowly-varying frequency induces AR only in the excited nonlinear actuator while the response of the coupled linear chain remains bounded.

It is expected that the difference in the systems dynamics is caused by their different resonant properties. In the system with a constant excitation frequency all oscillators are captured into resonance. The nonlinear oscillator remains captured into resonance due to an increase of the amplitude compensating the change of stiffness, while the partial frequency of the linear oscillators is always close to the excitation frequency. However, if the forcing frequency slowly increases, AR in the nonlinear oscillator is still sustained by the growth of the amplitude but the linear oscillators escape from resonance capture.

It is important to note that escape from resonance does not directly prevent further increase of energy in the linear oscillators. The linear chain is actually driven by the gradually increasing coupling response providing increasing energy transfer from the excited actuator to the coupled chain. Hence the system dynamics depends on the relationship between the growth of incoming energy and the loss of energy due to escape from resonance. Theoretical and numerical results presented in Sec. III prove that energy transfer in the system under consideration is insufficient to provide growing energy of the linear chain during escape from resonance. However, a numerical example demonstrates that an additional slow change of the actuator frequency may enhance energy transfer and make it sufficient to sustain growing oscillations of the coupled oscillator. Section IV contains a brief summary and conclusions.

**II. Autoresonance in chains with periodic excitations of constant frequency**

We consider a chain of *n* linear oscillators weakly coupled to a nonlinear actuator driven by external periodic forcing. The actuator represents the Duffing oscillator with slowly-decreasing linear stiffness. The equations of the uniaxial motion of the array are given by

$$m_0 \frac{d^2 u_0}{dt^2} + C(t)u_0 + \gamma u_0^3 + c_{0,1}(u_0 - u_1) = A\cos\omega t,$$
$$m_k \frac{d^2 u_k}{dt^2} + c_k u_k + c_{k,k-1}(u_k - u_{k-1}) + c_{k,k+1}(u_k - u_{k+1}) = 0, \ 1 \leq k \leq n-1, \quad (2.1)$$
$$m_n \frac{d^2 u_n}{dt^2} + c_n u_n + c_{n,n-1}(u_n - u_{n-1}) = 0.$$

In (2.1), $u_k$ represents the displacement of the *k-th* oscillator from the equilibrium position, $m_k$ is its mass; the coefficients $c_k$ and $c_{k,k+1} = c_{k+1,k}$ are the linear stiffness constants of the *k-th* oscillator and linear coupling between the *k-th* and *(k +1)-th* oscillators, respectively; $C(t) = c_0 - (\kappa_1 + \kappa_2 t)$ is the time-dependent linear stiffness of the Duffing oscillator, $\kappa_{1,2} > 0$; the parameters *A* and $\omega$ denote the amplitude and the frequency of the periodic force. The system is initially at rest, i.e., $u_k = 0$, $du_k/dt = 0$ at $t = 0$. It will be shown that a proper choice of linear stiffness $C(t)$ may serve as a tool to excite AR in the entire array.

The weakly coupled weakly nonlinear system (2.1) can be analyzed asymptotically. The small parameter $\varepsilon$ is defined by the equality $2\varepsilon = c_{1,0}/c_1 \ll 1$. Taking into account resonance properties of the system and assuming weak coupling and weak nonlinearity, we redefine the system parameters as follows:

$$c_k/m_k = \omega^2,\ \tau_0 = \omega t,\ \tau_1 = \varepsilon\tau_0,\ \kappa_1/c_0 = 2\varepsilon s,\ \kappa_2/c_0 = 2\varepsilon^2 b\omega, \quad (2.2)$$

$$\gamma/c_0 = 8\varepsilon\alpha,\ c_{k,k+1}/c_k = 2\varepsilon\lambda_{k,k+1},\ A = \varepsilon m\omega^2 F.$$

Equations (2.1) can now be rewritten as

$$\frac{d^2 u_0}{d\tau_0^2} + (1 - 2\varepsilon\zeta(\tau_1))u_0 + 2\varepsilon\lambda_0(u_0 - u_1) + 8\varepsilon\alpha u_0^3 = 2\varepsilon F\sin\tau_0,$$
$$\frac{d^2 u_k}{d\tau_0^2} + u_k + 2\varepsilon\lambda_{k,k-1}(u_k - u_{k-1}) + 2\varepsilon\lambda_{k,k+1}(u_k - u_{k+1}) = 0, 1 \leq k \leq n, \quad (2.3)$$

where $\zeta(\tau_1) = s + b\tau_1$. By definition, $\lambda_{0,1} = \lambda_0$, $\lambda_{1,0} = 1$, $\lambda_{k,k+1} \neq \lambda_{k+1,k}$ but $\lambda_{n+1,n} = \lambda_{n,n+1} = 0$. In analogy to a single oscillator [15], system (2.3) can be asymptotically solved with the help of the multiple scales method [16]. To this end, we introduce the complex-valued amplitudes

$$Y_k = (v_k + iu_k)e^{-i\tau_0},\ Y_k(0) = 0. \quad (2.4)$$

Inserting (2.4) into (2.3), we obtain the equations

$$\frac{dY_0}{d\tau_0} = -i\varepsilon[\zeta(\tau_1) - 3\alpha|Y_0|^2]Y_0 + i\varepsilon[\lambda_0(Y_0 - Y_1) - F] + i\varepsilon G_0(\tau, Y, Y^*),$$
$$\frac{dY_k}{d\tau_0} = i\varepsilon[\lambda_{k,k-1}(Y_k - Y_{k-1}) + \lambda_{k,k+1}(Y_k - Y_{k+1})] + i\varepsilon G_k(Y^*)e^{-2i\tau_0}, 1 \leq k \leq n, \quad (2.5)$$

where the term $G_0(\tau, Y, Y^*)$ involves the sum of harmonics with coefficients depending on $Y_1$, $Y_0$ and their complex conjugates, the term $G_k(Y^*)$ depends on $Y_k^*$, $Y_{k\pm1}^*$. It can be easily shown that an explicit form of the functions $G_0$ and $G_k$ is unimportant for further analysis.

It follows from (2.5) that the leading-order term in the asymptotic representation of $Y_k$ is independent of the fast time $\tau_0$. Hence, the following asymptotic expansion can be introduced

$$Y_k(\tau_0, \tau_1, \varepsilon) = y_k(\tau_1) + \varepsilon y_k^{(1)}(\tau_0, \tau_1) + \ldots, \quad (2.6)$$

Introducing rescaling transformations

$$\tau = s\tau_1, \ \Lambda = (s/3\alpha)^{1/2}, \ \psi_k = y_k/\Lambda, \qquad (2.7)$$

$$\mu_{k,k+1} = \lambda_{k,k+1}/s, \ f = F/s\Lambda, \ \beta = b/s^2, \ \zeta_0(\tau) = 1 + \beta\tau.$$

and then inserting (2.6), (2.7) into (2.5) and eliminating non-oscillating terms from the resulting equations, we obtain the following dimensionless equations for the amplitudes $\psi_k$:

$$\frac{d\psi_0}{d\tau} = -i[\zeta_0(\tau) - |\psi_0|^2]\psi_0 + i\mu_0(\psi_0 - \psi_1) - if.$$
$$\frac{d\psi_k}{d\tau} = i[\mu_{k,k-1}(\psi_k - \psi_{k-1}) + \mu_{k,k+1}(\psi_k - \psi_{k+1})], \qquad (2.8)$$

with initial conditions $\psi_0(0) = \psi_k(0) = 0$, $k = 1, \ldots, n$. The detailed derivation of the leading-order equations in similar systems can be found in earlier works [17, 18].

The real-valued amplitudes and phases of oscillations are defined as $a_r = |\psi_r|$, $\Delta_r = \arg\psi_r$. It now follows from (2.4), (2.7) that the leading-order approximation to the solution of Eqs. (2.3) is given by

$$u_k(\tau_0, \tau_1) = \Lambda a_k(s\tau_1)\sin(\tau_0 + \Delta_k(s\tau_1)), \ k = 0, 1, \ldots, n.$$

In the case of the light attachment such that $m_1 \ll m_0$, $m_1 = \varepsilon\delta_1 m_0$, $\delta_1 = O(1)$, one obtains $\mu_0 = O(\varepsilon)$. This implies the negligible influence of the attachment on the actuator in the main approximation. The leading-order approximations are governed by the truncated system

$$\frac{d\psi_0^{(0)}}{d\tau} = -i[\zeta_0(\tau_1) - 3\alpha|\psi_0^{(0)}|^2]\psi_0^{(0)} - if,$$
$$\frac{d\psi_k^{(0)}}{d\tau} = i[\mu_{k,k-1}(\psi_k^{(0)} - \psi_{k-1}^{(0)}) + \mu_{k,k+1}(\psi_k^{(0)} - \psi_{k+1}^{(0)})], \ 1 \le k \le n, \qquad (2.9)$$

in which the equation of the nonlinear oscillator can be considered separately. Corresponding approximations of the real-valued amplitudes and phases of oscillations are defined as $a_k^{(0)} = |\psi_k^{(0)}|$, $\Delta_k = \arg\psi_k^{(0)}$. The effect of weak coupling between the oscillators may be considered using the same iteration procedure as in the study of quasi-linear tunneling [17]. Note that the assumption $m_1 < m_0$ is made because it renders the equations asymptotically tractable.

Numerical simulations confirm that the dynamical characteristics analytically obtained for truncated system (2.9) hold true for a wider range of parameters.

We now briefly recall main results required for further analysis. It was shown [15] that AR in the Duffing oscillator may occur at $f > f_1 = \sqrt{2/27} \approx 0.272$, while the values $f < f_1$ corresponds to bounded oscillations at any rate $\beta$. In the domain $f > f_1$ the Duffing oscillator admits AR if $\beta < \beta^*$ and bounded oscillations (saturation) if $\beta > \beta^*$. The critical rate is defined as $\beta^* = [(f/f_1)^{2/3} - 1]/T^*$, where $\tau = T^*$ corresponds to the first minimum of the phase $\Delta_0(\tau)$ in the time-independent Duffing oscillator ($\beta = 0$). The value $T^*$ was also found both numerically and analytically. This paper is focused on the case $f > f_1$, $\beta < \beta^*$ corresponding to AR in a single Duffing oscillator.

As in a single oscillator (e.g., [19]), the solution $\psi_0(\tau)$ is interpreted as small fast fluctuations $\tilde{\psi}_0(\tau)$ near the quasi-steady state $\overline{\psi}_0(\tau)$, i.e., $\psi_0(\tau) = \overline{\psi}_0(\tau) + \tilde{\psi}_0(\tau)$. If we recall that $\overline{\psi}_0$ is calculated as a stationary point of the system with "frozen" value of $\zeta_0$ and assume $\mu_0 = O(\varepsilon)$, we obtain the following equation for $\overline{\psi}_0(\tau)$:

$$(\zeta_0 - |\overline{\psi}_0|^2)\overline{\psi}_0 = -f . \qquad (2.10)$$

The absolute value $\overline{a}_0(\tau) = |\overline{\psi}_0(\tau)|$ may be interpreted as the backbone curve. The solution of the cubic equation at $|f/2\zeta_0| \ll 1$ yields the following approximation

$$\overline{a}_0 \approx \sqrt{\zeta_0} \to \sqrt{\beta\tau}, \tau \to \infty \qquad (2.11)$$

with the phase $\overline{\Delta}_0 = \arg\overline{\psi}_0 = 0$. Note that equality (2.10) is derived from system (2.9), and the solution of (2.10) should be formally denoted as $\overline{\psi}_0^{(0)}(\tau)$. Close proximity of the functions $|\psi_0^{(0)}(\tau)|$ and $|\psi_0(\tau)|$ (Fig. 1(*a*)) allows omitting the upper index and considering (2.11) as an approximate backbone curve for the actuator in the entire system.

Once the solution $\overline{\psi}_0(\tau)$ is known, then, in analogy to the conservative case [20] asymptotic approximations for fluctuations $\tilde{\psi}_0(\tau)$ can be computed by linearizing the initial equation near the quasi-steady state $\overline{\psi}_0$.

In order to calculate responses $\psi_r(\tau)$, we rewrite the linear part of (2.8) in the form

$$\frac{d\Psi}{d\tau} + iM\Psi = -i\mu_{1,0}R\psi_0(\tau), \Psi(0) = 0. \qquad (2.12)$$

where $\Psi = (\psi_1,\ldots,\psi_n)^T$, $R = (1, 0,\ldots,0)^T$, $M$ corresponds to the matrix of the coefficients of the linear part of (2.8). It now follows that

$$\Psi(\tau) = -i\mu_{1,0}e^{-iM\tau}\int_0^\tau e^{iMs}R[\overline{\psi}_0(s) + \tilde{\psi}_0(s)]ds. \qquad (2.13)$$

Since the effect of small fast fluctuations $\tilde{\psi}_0(\tau)$ on the value of integral (2.13) is negligible compared to the contribution of the slow component $\overline{\psi}_0$, an analytical approximation for the solution $\Psi(\tau)$ is given by

$$\Psi(\tau) \approx -i\mu_{1,0}e^{-iM\tau}J(\tau), J(\tau) = \int_0^\tau e^{iMs}\overline{\psi}_0(s)dsR. \qquad (2.14)$$

where $\overline{\psi}_0(\tau) = \sqrt{1+\beta\tau}$. Integration by parts gives

$$J(s) = -iM^{-1}[e^{iM\tau}\overline{\psi}_0(\tau) - I\overline{\psi}_0(0)]R - \Phi(\tau),$$

$$\Phi(\tau) = \int_0^\tau e^{iMs}(d\overline{\psi}_0/ds)dsR,$$

where $d\overline{\psi}_0/ds = \beta/[2\sqrt{1+\beta s}]$. It is easy to deduce that each component $\phi_k$ of the vector $\Phi$ takes the form $\phi_k(\tau) = \sqrt{\beta}\Sigma_k(\tau)$, where $\Sigma_k(\tau)$ is the sum of the Fresnel integrals [21]. Hence $|\Sigma_k(\tau)| < C_k$, and

$$\Psi(\tau) = \overline{\Psi}(\tau) + \tilde{\Psi}(\tau) + O(\sqrt{\beta}), \qquad (2.15)$$

$$\overline{\Psi}(\tau) = -i\mu_{1,0}M^{-1}R\overline{\psi}_0(\tau), \tilde{\Psi}(\tau) = -e^{-iM\tau}\overline{\Psi}(0).$$

This implies that the response of each oscillator can be expressed as superposition of slow and fast components, i.e., $\psi_k(\tau) = \overline{\psi}_k(\tau) + \tilde{\psi}_k(\tau)$, where $\overline{\psi}_k(\tau)$ and $\tilde{\psi}_k(\tau)$ denote the quasi-steady state and fast fluctuations, respectively. It now follows from (2.8), (2.15) that $\overline{\psi}_k(\tau) = \overline{\psi}_0(\tau)$, i.e., coupled oscillators also exhibit AR. The fast-oscillating components $\tilde{\psi}_k(\tau)$ are given by

$$\tilde{\psi}_k(\tau) = \sum_{r=1}^{n} a_{kr} e^{i(\omega_r \tau + \gamma_{kr})}, k = 1,...,n, \qquad (2.16)$$

where $i\omega_r$ are the roots of the characteristic polynomial of system (2.12), $a_{kr}$ and $\gamma_{kr}$ are the real-valued amplitude and the phase of the *r-th* harmonic.

As an example, we analyse the dynamics of the chain of 3 equal oscillators linearly coupled to the forced Duffing oscillator. In this case Eqs (2.8) take the form

$$\begin{aligned}
&\frac{d\psi_0}{d\tau} + i[\zeta(\tau_1) - |\psi_0|^2]\psi_0 - i\mu_0(\psi_0 - \psi_1) = -if, \\
&\frac{d\psi_1}{d\tau} - i\mu(2\psi_1 - \psi_2) = -i\mu\psi_0, \\
&\frac{d\psi_2}{d\tau} - i\mu(2\psi_2 - \psi_1 - \psi_3) = 0, \\
&\frac{d\psi_3}{d\tau} - i\mu(\psi_3 - \psi_2) = 0.
\end{aligned} \qquad (2.17)$$

Figure 1 depicts the amplitudes of oscillations in system (2.17) with the following parameters:

$$\beta = 0.05, f = 0.34, \mu_0 = 0.015, \mu = 0.1 \qquad (2.18)$$

It is easy to check that the single nonlinear oscillator with parameters (2.18) admits autoresonance (see [15]) and the truncated system ($\mu_0 = 0$) correctly approximates the dynamics of the full system.

It is seen in Fig. 1 that the amplitude of the Duffing oscillator $a_0(\tau)$ represents the superposition of small fast oscillations to the monotonically increasing backbone curve $\overline{a}_0(\tau)$, while all other amplitudes $a_k(\tau)$ involve low-frequency oscillatory components.

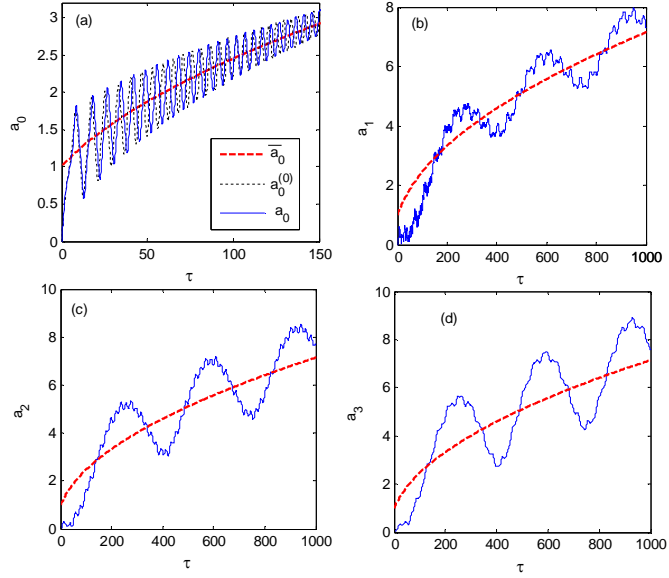

FIG. 1. Amplitudes of autoresonance oscillations in the chain (2.17); dashed lines correspond to the backbone curves $\bar{a}_r$; dotted lines in Fig. 1(a) depict the amplitude of the nonlinear oscillator in the truncated system

From Fig. 1(a) it is seen that the amplitude $a_0$ of the actuator in the full system (2.17) is close to the amplitude $a_0^{(0)}$ of the nonlinear oscillator in the truncated system. Since these solutions are in fairly good agreement, the above-described approximate procedure can be applied to system (2.17). In other words, the nonlinear equation can be formally solved at $\mu_0 = 0$ and the obtained approximation is then substituted in the linear part of (2.17).

The characteristic polynomial of the linear subsystem is expressed as $D(s) = (s - i\mu)(s - 2i\mu)^2 + \mu^2(2s - 3i\mu)$; the roots of the characteristic equation $D(s) = 0$ take the form $s_k = i\omega_k$, where $\omega_k = \mu\alpha_k$, $\alpha_1 = 0.2$, $\alpha_2 = 1.56$, $\alpha_3 = 3.25$. The period of the dominant low-frequency harmonic $T_1 = 2\pi/\omega_1 = 314$ is obviously close to the exact (numerical) value $T \approx 350$ (see Figs. 1(b) - 1(d)); the difference between the analytical and numerical results is about 10%. The initial conditions $\tilde{\psi}_r(0) = -1$ define the following amplitudes of the dominant low-frequency harmonics $\omega_1$ in the $r$-th oscillator ($r = 1, 2, 3$): $a_{11} = 0.63$; $a_{21} = 0.97$, $a_{31} = 1.22$. It is easy to verify both analytically and numerically that the higher-frequency amplitudes $a_{r2}$ and $a_{r3}$ satisfy the conditions $a_{r2} \ll a_{r1}$, $a_{r3} \ll a_{r1}$.

## III. Energy localization and transport in chains excited by forcing with slowly-varying frequency

In this section we briefly analyse the dynamics of a system, wherein the nonlinear actuator with constant parameters is subjected to external forcing with slowly changing frequency. It is important to note that slow variation of a frequency of the external force is a common way to excite AR in a single oscillator (see, e.g., a number of examples in [4]). We demonstrate that this kind of frequency control cannot be directly applicable for a multi-degree of freedom system.

We consider the dimensionless equations, wherein the group of the linear equations coincide with the same group in (2.3) but the equation of the actuator takes the form

$$\frac{d^2 u_0}{d\tau_0^2} + u_0 + 2\varepsilon\lambda_0(u_0 - u_1) + 8\varepsilon a u_0^3 = 2\varepsilon F \sin(\tau_0 + \theta(\tau_1)),$$

$$\frac{d\theta}{d\tau_1} = \zeta(\tau_1).$$
(3.1)

Transformations (2.4) – (2.7) yield the equations for the dimensionless complex amplitudes $\psi_r$ similar to (2.8)

$$\frac{d\psi_0}{d\tau} = i|\psi_0|^2 \psi_0 + i\mu_0(\psi_0 - \psi_1) - ife^{i\theta_0(\tau)},$$

$$\frac{d\psi_k}{d\tau} = i[\mu_{k,k-1}(\psi_k - \psi_{k-1}) + \mu_{k,k+1}(\psi_k - \psi_{k+1})],$$
(3.2)

$$\frac{d\theta_0}{d\tau} = \zeta_0(\tau),$$

with initial conditions $\psi_0(0) = \psi_k(0) = 0$, $k = 1, \ldots, n$. In the next step, the change of variables $\phi_k(\tau) = \psi_k(\tau)e^{-i\theta_0(\tau)}$ transforms Eqs. (3.2) into the system with slowly-varying coefficients and a constant right-hand side:

$$\frac{d\phi_0}{d\tau} + i[\zeta_0(\tau) - |\phi_0|^2]\phi_0 - i\mu_0(\phi_0 - \phi_1) = -if,$$

$$\frac{d\phi_k}{d\tau} + i\zeta_0(\tau)\phi_k - i[\mu_{k,k-1}(\phi_k - \phi_{k-1}) + \mu_{k,k+1}(\phi_k - \phi_{k+1})] = 0,$$
(3.3)

where $\phi_0(0) = \phi_k(0) = 0$, $k = 1, \ldots, n$. It is important to note that slow detuning $\zeta_0(\tau)$ is now included both in the nonlinear equation as well as in the linear system. As in the previous section, the solution $\phi_0(\tau)$ is represented as $\phi_0(\tau) = \bar{\phi}_0(\tau) + \tilde{\phi}_0(\tau)$, where $\bar{\phi}_0(\tau)$ and $\tilde{\phi}_0(\tau)$ denote the quasi-steady state of the nonlinear oscillator and fast fluctuations near $\bar{\phi}_0(\tau)$, respectively. Under the condition $\mu_0 \sim O(\varepsilon)$, the function $\bar{\phi}_0$ satisfies the equation similar to (2.10), and $\bar{\phi}_0 \approx \sqrt{\zeta_0}$ at large times. If the solution $\phi_0(\tau)$ is known, all other variables can be calculated from (3.3). In analogy with (2.8), (2.12), the linear part of (3.3) is rewritten as

$$\frac{d\Phi}{d\tau} + iM_1(\tau)\Phi = -i\mu_{1,0}R\phi_0(\tau), \Phi(0) = 0. \tag{3.4}$$

where $\Phi = (\phi_1, \ldots, \phi_n)^T$, $R = (1, 0, \ldots, 0)^T$, $M_1(\tau) = \zeta_0(\tau)I + M$, $I$ is the unit matrix, $M$ is the matrix of system (2.12). In analogy to (2.14), we obtain

$$\Phi(\tau) \approx -i\mu_{1,0}e^{-iM_2(\tau)}K(\tau),$$
$$K(\tau) = \int_0^\tau e^{iM_2(s)}\bar{\phi}_0(s)ds R, \quad M_2(\tau) = \int_0^\tau M_1(s)ds. \tag{3.5}$$

We now show that each component of the vector $\Phi(\tau)$ is bounded, i.e. $|\phi_r(\tau)| < c_r < \infty$, $\tau \geq 0$. For brevity, the model of two coupled oscillators is considered. The slow dynamics of this system is described by the equations

$$\frac{d\phi_0}{d\tau} + i[\zeta_0(\tau_1) - |\phi_0|^2]\phi_0 - i\mu_0(\phi_0 - \phi_1) = -if,$$
$$\frac{d\phi_1}{d\tau} + i[\zeta_0(\tau) - \mu]\phi_1 = -i\mu\phi_0. \tag{3.6}$$

with zero initial conditions. The change of variables $S(\tau) = (1+\beta\tau)^2$ and simple transformations reduce the solution $\phi_1(\tau)$ to the form

$$\phi_1(\tau) \approx -i\frac{\mu}{2\beta}e^{-iS(\tau)/2\beta}K(\tau), K(\tau) = K_0(\tau) - K_0(1),$$
$$K_0(\tau) = \int_0^{S(\tau)} e^{iz/2\beta}z^{-1/4}dz. \tag{3.7}$$

Although the expression for $K_0(\tau)$ cannot be analytically defined, the limiting value $K_0(\infty)$ can be explicitly evaluated, and equals $K_0(\infty) = (2\beta)^{4/3}\Gamma(\sfrac{3}{4})e^{3i\pi/8}$, where $\Gamma$ is the gamma function [21]. Hence $a_1(\tau) = |\phi_1(\tau)| \to \mu(2\beta)^{1/3}\Gamma(\sfrac{3}{4})$ as $\tau \to \infty$. Therefore, energy of the drive mainly remains localized on the excited oscillator but the rest of energy transferred to the linear oscillator suffices to sustain motion with bounded non-decaying amplitude. Similar reasoning applied to the multi-dimensional array allows concluding that the response of each linear oscillator can also be represented as the sum of harmonics with bounded amplitudes.

As an example, we consider a four-dimensional chain. It follows from (3.3) that the equations for the dimensionless amplitudes $\phi_r$ are given by

$$\begin{aligned}
\frac{d\phi_3}{d\tau} + i\zeta_0(\tau)\phi_3 - i\mu(\phi_3 - \phi_2) &= 0, \\
\frac{d\phi_2}{d\tau} + i\zeta_0(\tau)\phi_2 - i\mu(2\phi_2 - \phi_1 - \phi_3) &= 0, \\
\frac{d\phi_1}{d\tau} + i\zeta_0(\tau)\phi_1 - i\mu(2\phi_1 - \phi_2) &= -i\mu\phi_0, \\
\frac{d\phi_0}{d\tau} + i[\zeta_0(\tau_1) - |\phi_0|^2]\phi_0 - i\mu_0(\phi_0 - \phi_1) &= -if.
\end{aligned} \qquad (3.8)$$

The amplitudes of oscillations $a_k(\tau) = |\phi_k(\tau)|$ for system (3.8) with parameters (2.18) are shown in Fig. 2. The obtained numerical results confirm a bounded response of all linear oscillators despite AR with permanently growing amplitude in the forced nonlinear oscillator.

As mentioned earlier in this paper, if the forcing frequency slowly increases, the linear oscillators escape from resonance. It is important to note that escape from resonance does not directly prevent the growth of energy, as the linear chain is actually driven by the coupling response with permanently increasing amplitude, and the system dynamics depends on the relationship between the growth of incoming energy and the loss of energy due to exit from resonance. It is seen in Fig. 2 that energy transfer from the excited oscillator is insufficient to render oscillations with increasing energy in the linear chain during escape from resonance.

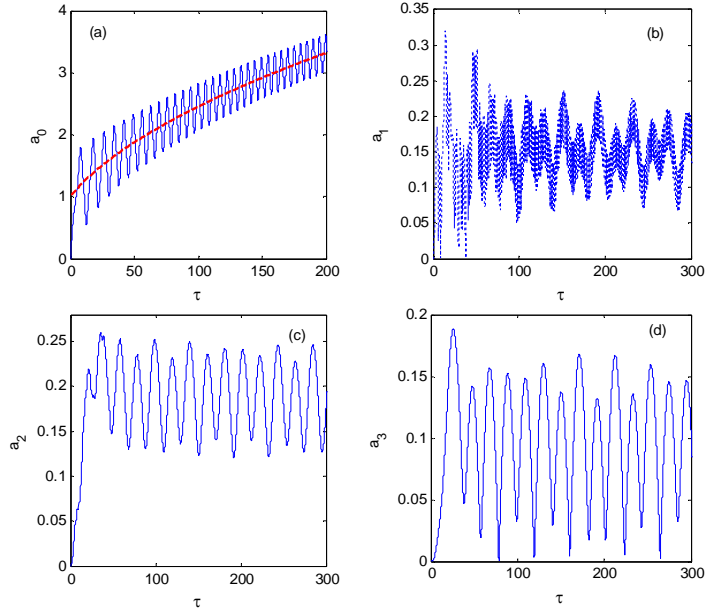

FIG. 2. Amplitudes of the actuator and the coupled oscillators in the chain (3.8)

As shown below, slow changes in both natural and excitation frequencies of the actuator may entail growing oscillations in the coupled linear oscillator. For brevity, we consider a 2DOF system. The dimensionless equations of the system take the form similar to (2.3)

$$\frac{d^2 u_0}{d\tau_0^2} + (1 - 2\varepsilon\xi(\tau_1))u_0 + 2\varepsilon\lambda_0(u_0 - u_1) + 8\varepsilon a u_0^3 = 2\varepsilon F \sin(\tau_0 + \theta(\tau_1)),$$
$$\frac{d^2 u_1}{d\tau_0^2} + u_1 + 2\varepsilon\lambda_{1,0}(u_1 - u_0) = 0, \qquad (3.9)$$
$$\frac{d\theta}{d\tau_1} = \zeta(\tau_1),$$

where $\tau_1 = \varepsilon\tau_0$, $\zeta(\tau_1) = s + b_1\tau_1$, $\xi(\tau_1) = b_3\tau_1^3$; all other parameters are defined in (2.1) - (2.3). Recall that AR may appear only the nonlinear oscillator if $\xi(\tau_1) = 0$.

Transformations (2.4)-(2.7) yield the equations for the dimensionless complex amplitudes $\psi_r$ similar to (2.8)

$$\frac{d\psi_0}{d\tau} = -i[\xi_1(\tau) - |\psi_0|^2]\psi_0 + i\mu_0(\psi_0 - \psi_1) - if e^{i\theta_0(\tau)},$$
$$\frac{d\psi_1}{d\tau} = i\mu_1(\psi_1 - \psi_0), \qquad (3.10)$$
$$\frac{d\theta_0}{d\tau} = \zeta_1(\tau_1),$$

where

$$\tau = s\tau_1,\ \zeta_1(\tau) = 1 + \beta_1\tau,\ \xi_1(\tau) = \beta_3\tau^3,\ \beta_1 = b_1/s^2,\ \beta_3 = b_3/s^4,$$

$$\mu_0 = \lambda_0/s,\ \mu_1 = \lambda_{1,0}/s,\ f = F/s\Lambda,\ \mu_0 = \lambda_0/s,\ \mu_1 = \lambda_{1,0}/s,\ f = F/s\Lambda.$$

Finally, the change of variables $\phi_k(\tau) = \psi_k(\tau)e^{-i\theta_0(\tau)}$ transforms (3.10) into the equations with a constant right-hand side and time-dependent coefficients

$$\begin{aligned}&\frac{d\phi_0}{d\tau} + i[\zeta_0(\tau) - |\phi_0|^2]\phi_0 - i\mu_0(\phi_0 - \phi_1) = -if,\ \phi_0(0) = 0,\\ &\frac{d\phi_1}{d\tau} + i\zeta_1(\tau)\phi_1 - i\mu_1(\phi_1 - \phi_0) = 0,\ \phi_1(0) = 0,\end{aligned} \qquad (3.11)$$

where $\zeta_0(\tau) = \zeta_1(\tau) + \xi_1(\tau)$. The amplitudes and the phases of oscillations are expressed as $a_k = |\phi_k|$ and $\Delta_k = \arg\phi_k$, respectively.

Figure 3 depicts the amplitudes of oscillations in the system with parameters

$$\beta_1 = 10^{-3},\ \beta_3 = 10^{-5}, f = 0.34,\ \mu_0 = 0.01,\ \mu_1 = 0.15. \qquad (3.12)$$

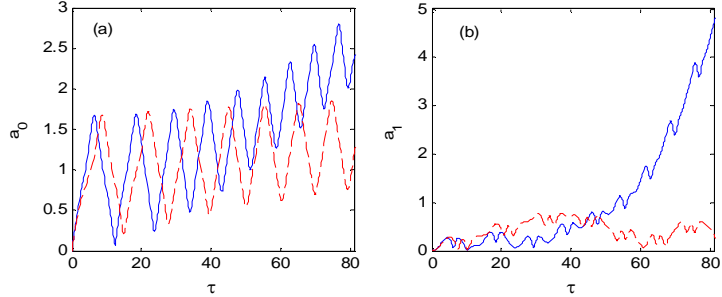

FIG. 3. Amplitudes of oscillations of the actuator (*a*) and the linear oscillator (*b*); solid lines corresponds to system (3.11) with parameters (3.12); dashed lines correspond to the time-independent actuator ($\beta_3 = 0$).

Hence an additional slow change of the actuator stiffness can increase the nonlinear response, thereby enhancing energy transfer and making it sufficient to sustain growing oscillations of the linear oscillator. Plots in Fig. 3 depict the amplitudes of both oscillators in the systems with and without additional detuning of the actuator frequency. A theoretical analysis of this model is omitted but the obtained numerical results motivate further analytical investigation of feasible energy transfer in the multi-dimensional arrays.

**IV. CONCLUSIONS**

It was shown in early works on particle acceleration that autoresonance could potentially serve as a mechanism to excite and control the required high-energy regime in a single oscillator. This principle was further employed in various fields of applied physics. However, the behavior of the multi-dimensional array can drastically differ from the dynamics of a single oscillator. This paper has investigated an array consisting of a time-independent linear oscillator chain weakly coupled with a forced nonlinear actuator (the Duffing oscillator). Two types of systems have been considered in details: in the system of the first type a harmonic excitation with constant frequency is applied to the Duffing oscillator with slowly time-decreasing linear stiffness; in the system of the second type the nonlinear oscillator with constant parameters is subjected to harmonic forcing with slowly increasing frequency. In both cases, stiffness of the linear oscillators and coupling remain constant, and the system is initially engaged in resonance. It has been shown both theoretically and numerically that in the first case AR occurs in all oscillators and provides the equal distribution of the mean energy between all oscillators in the chain. In the second case, AR occurs only in the excited actuator as the energy transfer from the excited oscillator is insufficient to cause motion with growing energy in the coupled chain. However, a simple example demonstrates that a proper slow change of the actuator frequency may enhance energy transfer and make it sufficient to sustain gradually increasing energy of the coupled oscillator. A careful analysis of feasible energy transport in multi-dimensional chains of a more complicated structure is left for further works.

**Acknowledgment**

Support for this work received from the Russian Foundation for Basic Research through the RFBR grant 14-01-00284 is gratefully acknowledged. The author would like to thank L.I. Manevitch for valuable discussion.